\documentclass[aps]{revtex4}

\usepackage{graphicx}
\usepackage{amsmath} 

\begin{document}

\newcommand{\rme}{{\rm e}}
\newcommand{\rmd}{{\rm d}}
\newcommand{\bl}{{l}}
\newcommand{\bk}{{k}}
\newcommand{\Kbp}{\bar{K}^{\prime}}
\newcommand{\bA}{{\bf A}}
\newcommand{\w}{\omega}
\newcommand{\D}{\Delta}
\newcommand{\Om}{\Omega}
\newcommand{\la}{\lambda}
\newcommand{\lab}{\bar{\lambda}}
\newcommand{\La}{\Lambda}
\newcommand{\e}{\epsilon}
\newcommand{\n}{\eta}
\newcommand{\nb}{\bar{\eta}}
\newcommand{\np}{\eta^{\prime}}
\newcommand{\nbp}{\bar{\eta}^{\prime}}
\newcommand{\psib}{\bar{\psi}}
\newcommand{\psip}{\psi^{\prime}}
\newcommand{\psibp}{\bar{\psi}^{\prime}}
\newcommand{\phib}{\bar{\phi}}
\newcommand{\phip}{\phi^{\prime}}
\newcommand{\phibp}{\bar{\phi}^{\prime}}
\newcommand{\z}{z}
\newcommand{\zb}{\bar{z}}
\newcommand{\zp}{z^{\prime}}
\newcommand{\zbp}{\bar{z}^{\prime}}
\newcommand{\f}{f}
\newcommand{\fb}{\bar{f}}
\newcommand{\fp}{f^{\prime}}
\newcommand{\fbp}{\bar{f}^{\prime}}
\newcommand{\g}{g}
\newcommand{\gb}{\bar{g}}
\newcommand{\gp}{g^{\prime}}
\newcommand{\gbp}{\bar{g}^{\prime}}
\newcommand{\F}{F}
\newcommand{\Fbp}{\bar{F}^{\prime}}
\newcommand{\G}{G}
\newcommand{\Gbp}{\bar{G}^{\prime}}
\newcommand{\Psif}{\Psi^f}
\newcommand{\Psifbp}{\bar{\Psi}^{\prime f}}
\newcommand{\Psig}{\Psi^g}
\newcommand{\Psigbp}{\bar{\Psi}^{\prime g}}
\newcommand{\Phif}{\Phi^f}
\newcommand{\Phifbp}{\bar{\Phi}^{\prime f}}
\newcommand{\Phig}{\Phi^g}
\newcommand{\Phigbp}{\bar{\Phi}^{\prime g}}
\newcommand{\dF}{\dot{F}}
\newcommand{\dFbp}{\dot{\bar{F}}^{\prime}}
\newcommand{\dG}{\dot{G}}
\newcommand{\dGbp}{\dot{\bar{G}}^{\prime}}
\newcommand{\dPsif}{\dot{\Psi}^f}
\newcommand{\dPsifbp}{\dot{\bar{\Psi}}^{\prime f}}
\newcommand{\dPsig}{\dot{\Psi}^g}
\newcommand{\dPsigbp}{\dot{\bar{\Psi}}^{\prime g}}
\newcommand{\dPhif}{\dot{\Phi}^f}
\newcommand{\dPhifbp}{\dot{\bar{\Phi}}^{\prime f}}
\newcommand{\dPhig}{\dot{\Phi}^g}
\newcommand{\dPhigbp}{\dot{\bar{\Phi}}^{\prime g}}
\newcommand{\definedas}{\stackrel{\text{def}}{=}}

\title{Non-Markovian qubit dynamics in a thermal field bath:\\ Relaxation, decoherence and entanglement}
\author{S.~Shresta} \thanks{Present Address: NIST, Atomic Physics Division, Gaithersburg, MD 20899-8423.
Email: sanjiv.shresta@nist.gov}
\author{C.~Anastopoulos} \thanks{Present Address: Department of Physics, University of Patras,
26500 Patras, Greece. Email: anastop@physics.upatras.gr }
\author{A.~Dragulescu} \thanks{Present Address: Constellation Energy Group, Baltimore, MD 21202. Email: adrian.dragulescu@constellation.com }
\author{B.~L.~Hu} \thanks{Email: hub@physics.umd.edu}
\affiliation{Department of Physics, University of Maryland,
College Park, Maryland 20742}
\date{\today}

\begin{abstract}
We study the non-Markovian dynamics of a qubit made up of a
two-level atom interacting with an electromagnetic field (EMF)
initially at finite temperature. Unlike most earlier studies where
the bath is assumed to be fixed, we study the coherent evolution
of the combined qubit-EMF system, thus allowing for the
back-action from the bath on the qubit and the qubit on the bath
in a self-consistent manner. In this way we can see the
development of quantum correlations and entanglement between the
system and its environment, and how that affects the decoherence
and relaxation of the system. We find non-exponential decay for
both the diagonal and non-diagonal matrix elements of the qubit's
reduced density matrix in the pointer basis. From the diagonal
elements we see the
qubit relaxes to thermal equilibrium with the bath. From the
non-diagonal elements, we see the decoherence rate beginning at
the usually predicted thermal rate, but changing to the zero
temperature decoherence rate as the qubit and bath become
entangled. These two rates are comparable, as was shown before in
the zero temperature case  [C. Anastopoulos and B. L. Hu, Phys.
Rev. A {\bf 62} (2000) 033821]. On the entanglement of a qubit
with the EMF under this type of resonant coupling we calculated, for
the qubit reduced density matrix,
the fidelity and the von~Neumann entropy, which
is a measure of the purity of the density matrix. The present more
accurate non-Markovian calculations predict lower loss of fidelity
and purity as compared with the Markovian results. Generally
speaking,  with the inclusion of quantum correlations between the
qubit and its environment, the non-Markovian processes tend to
slow down the drive of the system to equilibrium, prolonging the
decoherence and better preserving the fidelity and purity of the
system.
\end{abstract}
\maketitle

\section{Introduction}
Interest in quantum entanglement has grown in recent years
motivated mainly by the attempt to understand and realize quantum
information processing.
An important aspect of quantum
entanglement, which is unavoidable in quantum information processing,
is the entanglement of a system with its environment. This issue is especially important to the feasibility of quantum computation, as the error threshold for fault tolerant error correction (without which quantum computation is impossible), is highly sensitive to the degree to which the environment becomes correlated with the qubits~\cite{Knill}.
In many realistic quantum computing schemes
the environment includes an electromagnetic field (EMF). Studies of entanglement between multi-partite systems
exist~\cite{AbdelAty02,Eisert02,Keyl03,Brennen03,Latorre04}, but few have
attempted to predict the effects
on a qubit from its entanglement with the
EMF~\cite{BarnesWarren99,GeaBanacloche02,vanEnk02,SilberfarbDeutsch04}. We have
addressed the relaxation and decoherence issues in various
contexts, from a two-level atom in a (zero-temperature) EMF~\cite{ABH} 
to moving atoms in a cavity~\cite{SH1,SH2}. Here we
continue this study for these two issues and the issue of
entanglement for a two-level atom in a finite temperature EMF.

In this paper we aim at addressing the issue of system-environment entanglement by carefully analyzing its effect on reduced system dynamics. Specifically, we study the
reduced dynamics of a single qubit interacting with an initially thermal
bath in the multimode Jaynes-Cummings model. The model is a well
studied example of open system dynamics, however,
prior analyses have focused on Markovian dynamics by assuming no
disturbance of the bath modes by the
qubit~\cite{carmichael,Louisell,WM,ctan,Scu,GardinerAndZoller}.
Although such a simplifying assumption does include
a level of back-action, it neglects entanglement that forms
between the qubit and bath modes during the evolution.
 Earlier predictions based on simplified
Markovian approximations are thus unsuitable for studies of such entanglement, and should be scrutinized carefully before being applied to quantum computation (see e.g.~\cite{AlickiHorodecki}
for a discussion of this issue bearing on error correction). More realistic physical conditions are better served by non-Markovian
treatments, which have begun to appear. However, many of them use approximations with limited short-time validity~\cite{KimNemesPizaBorges96,BreuerMaPetruccione02,EspositoGaspard03,TolkunovPrivman04} or unclear physical meaning such as in non-Markovian stochastic Schroedinger equations~\cite{GambettaWiseman,Yu}.

Our approach includes bath dynamics as well as qubit dynamics, and the
quantum correlations between them. Analytic expressions are derived
for the qubit reduced density matrix elements at low temperature, from which the fidelity (defined as the overlap between the initial and evolved qubit state) and von~Neumann entropy are computed. Inspection of the matrix elements themselves show slowed (sub-exponential) decoherence for the off-diagonal elements and slowed relaxation to thermal asymptotes for the diagonal matrix elements, as compared to Markovian predictions. Analysis of the fidelity and von~Neumann entropy similarly show slowed loss of fidelity and purity in the case of non-Markovian dynamics over Markovian dynamics. The overall picture which develops is of increased preservation of coherence in non-Markovian dynamics.

Following successive degradation from an exact solution to a
Markovian description of system-bath interactions, three distinct
approximations are usually invoked. They are 1) the $2^{nd}$ order
Born approximation, 2) the $1^{st}$ Markov
approximation~\cite{GardinerAndZoller}, and 3) the assumption of a
bath which is unaffected by its interaction with the system. 1)
The $2^{nd}$ order Born approximation is an approximation in the
strength of the coupling constant, and applying it neglects terms
of higher than $2^{nd}$ order in the coupling~\cite{Haake}. 2) The
$1^{st}$ Markov approximation is an approximation in the
back-action correlation time. It is a Markov approximation because
it assumes that the back-action of the system onto itself through
the bath at time $t$ will depend only on the state of the combined
system-bath at time $t$, and not on the past history. It is called
the $1^{st}$ Markov approximation because although it depends only
on the state of the system at time $t$, it depends on the state of
the bath as well as the system (through the bath averages), thus
including the bath dynamics~\cite{GardinerAndZoller}. 3) The last
of the above three approximations is the assumption of a bath
state which is fixed for all time. That assumption expressly
excludes any dynamical evolution of the bath.

In the usual derivation of the Markovian master equation from the
Schr\"odinger equation for the system-bath density operator, all
three approximations are made. In contrast, the
Heisenberg-Langevin approaches make only the $1^{st}$ Markov
approximation. However, for spin-boson models such approaches have
focused on the reduced qubit dynamics in strictly vacuum EMF,
although in the presence of a classical source (e.g.~resonance
florescence). The resulting equations for the qubit degrees of
freedom are called the Bloch-Langevin equations~\cite{ctan}. The
Schr\"odinger-master equation can be derived from the
Heisenberg-Langevin equation after a perturbative expansion which
imposes the first and third approximations from the
above~\cite{vanKampen}. For a comparison of these approximations
see Ref.~\cite{ShrestaPhD}. Our path integral approach to the
reduced system dynamics uses only the first of the above three
approximations by allowing the combined system-bath to evolve
coherently throughout the interaction period. Only at the end of
all coherent evolution will the bath variables be traced out to
yield the reduced system evolution.

The approach we take is straightforward, although the actual
implementation includes some nonstandard techniques involving
Grassmann path integrals. In Section~II the Hamiltonian and other
important aspects of the model, including the coherent state
represenation, are reviewed. The transition amplitude is derived
in Section~IIIA, utilizing the coherent state representation for
the bosonic degrees of freedom and Grassmann states for the qubit
degrees of freedom, following~\cite{ABH}. Doing so will involve a
recursive computation which exploits the semigroup property of the
transition matrix (similar to a technique used in
Ref.~\cite{GelfandAndYaglom}). After evaluating the transition amplitudes in
an intermediate form, we combine the forward and backward versions
by tracing over the final bosonic coherent states to construct the
reduced propagator in Section~IIIB. An initial thermal state for
the oscillator bath is then introduced and the reduced dynamics of
the qubit are calculated in Section~IIIC. Section~IV gives
discussion and further analysis of the results.

\section{Model}
\subsection{Hamiltonian}
The model used for atom-field interactions is the standard
multi-mode Jaynes-Cummings model of a two-level system interacting
with a harmonic oscillator bath. The total Hamiltonian under the
dipole, rotating wave and two-level approximations is given
by (e.g.~Appendix~A of~\cite{ABH})
\begin{eqnarray}\label{hamiltonian1} H = \hbar \omega_o S_z +
\hbar \sum_\bk \left[ \omega_{\bk} b_{\bk}^{\dagger} b_{\bk} +
\left(\la_{\bk}  S_+ b_{\bk}  +\bar{\la}_{\bk}  S_
-b_{\bk}^{\dagger} \right) \right]
\end{eqnarray}
where $\hat{b}_{\bk}^{\dagger},\hat{b}_{\bk}$ are the creation and
annihilation operators for the $\bk^{th}$ bath mode with frequency
$\omega_{\bk}$, and $ \hbar\omega_o$ is the energy separation
between the two levels. The operators $S_z$, $S_+$, and $S_-$ are
the qubit operators for z-projection, spin-up, and spin-down,
respectively. The couplings, $\la_k$ and $\lab_k$, have absorbed a
dependence on the spectral density of the bath~\cite{ABH}.

\subsection{Coherent States}
Coherent states are defined as any set of states generated by the
exponentiated operation of a creation operator and a suitable
label on a chosen fiducial state~\cite{ohnuki,perelomov}:
\begin{eqnarray}
\label{emcoherentstates} |z_\bk \rangle &=& \exp(z_\bk b_\bk^\dagger) |0_\bk \rangle \\
\label{gmcoherentstates} |\n \rangle &=& \exp(\n S_+) |0\rangle
\end{eqnarray}
In the case of the bosonic coherent states, defined in
Eq.~(\ref{emcoherentstates}), the label $z_\bk$ is a complex
number, and in the case of the Grassmann coherent states, defined
in Eq.~(\ref{gmcoherentstates}), the label $\n$ is an
anti-commuting number. The chosen fiducial states are the EMF
vacuum and the lower state of the two-level system, respectively. A
state of the combined atom-field system can be expanded in a
direct product coherent state basis,
\begin{equation}
|\{z_\bk\},\n  \rangle =|\{z_\bk\}\rangle \otimes |\n  \rangle,
\end{equation}
in which the bosonic coherent state basis, $|z_\bk\rangle $, is
used to represent the EMF, and a Grassmann coherent state basis,
$|\n\rangle$, is used to represent the two-level internal degrees of
freedom of the atom.

Grassmann and bosonic coherent states share well known properties
of general coherent states, such as being non-orthogonal and
eigenstates of the annihilator,
\begin{eqnarray}
\label{non-orthogonality}\langle \zb_\bk|\zp_\bk\rangle &=&
\exp(\zb_\bk \zp_\bk) \mbox{ }\mbox{ }\mbox{ }\mbox{ }\mbox{
}\langle \nb|\np\rangle = \exp(\nb \np)
\\
\label{annihilatoreigen}b_\bk |z_\bk\rangle &=& z_\bk
|z_\bk\rangle \mbox{ }\mbox{ }\mbox{ }\mbox{ }\mbox{ }\mbox{
}\mbox{ }\mbox{ }\mbox{ }\mbox{ }\mbox{ } S_- |\n\rangle = \n
|\n\rangle,\mbox{ }\mbox{ }\mbox{ }\mbox{ }
\end{eqnarray}
where the overbar denotes conjugation.
Despite their non-orthogonality, both types of coherent states are
(over-)complete sets and have a resolution of unity,
\begin{eqnarray}
\label{completeness} 1 = \int {\rmd}\mu(z_\bk)
|z_\bk\rangle\langle\zb_\bk| =\int {\rmd}\mu(\n)
|\n\rangle\langle\nb|
\end{eqnarray}
with the measures
\begin{eqnarray}
{\rmd}\mu(z_\bk) &=& \exp(-\zb_\bk z_\bk) \rmd\zb_k \rmd z_k\\
{\rmd}\mu(\n) &=& \exp(-\nb\n) \rmd\nb \rmd\n.
\end{eqnarray}
That these measures are exponential functions makes the coherent
states a particularly suitable representation for transition
amplitudes written as path integrals. For convenience the short hand notation
\begin{equation}
\prod_k\rmd\mu(\z_\bk)=\rmd\mu(\{\z_\bk\})
\end{equation}
is defined for the product of the measure of different mode coherent states.

In the bosonic and Grassmann coherent states, the Q-representation
of the Hamiltonian, Eq.~(\ref{hamiltonian1}), is
\begin{eqnarray}\label{hamiltonian2}
\langle \nb, \{\zb_{k}\} | H  | \np, \{\zp_{k}\} \rangle = \hbar \omega_o \nb \np + \hbar \sum_\bk \left[ \omega_{\bk}
\zb_\bk \zp_\bk + \left(\la_{\bk} \nb \zp_\bk  +\lab_{\bk}  \zb_\bk
\np \right) \right], \nonumber
\end{eqnarray}
in which the replacement $S_z\rightarrow S_+ S_-$,  correct up to
an additive constant, was made. The Q-representation Hamiltonian
will participate prominently in the path integrals of the next
section.

\section{Approach}
\subsection{Transition Amplitude}
Here we construct and evaluate the transition amplitude
$K(t_f,t_i)$ of coherent states from an initial time, $t_i=0$, to coherent states at a final time, $t_f=t$,
\begin{equation}
K(t,0) = \langle \nb_f, \{\zb_{fk}\} ;t| U(t,0) | \n_i, \{\z_{ik}\} ;0\rangle .
\end{equation}
with $U(t,0)$ being the time evolution operator,
\begin{equation}
U(t,0) = e^{-\frac{i}{\hbar} \int_0^t H {\rm ds}}.
\end{equation}
Following the path integral methodology, we partition the interval
$[0,t]$ into a large number ($N$) of time steps, such that
$t=N\e$. The path integral is then calculated as a discrete
functional. Doing so, the n-step transition amplitude can be
written in a general form,
\begin{equation} \label{nstepprop}
K(n\e,0) = \exp \bigg\{ \nb_n \psi_n +\sum_k \zb_{nk} f_{nk}
+\sum_k \nb_n g_{nk} +\sum_k \zb_{nk} \phi_{nk} \bigg\}.
\end{equation}
By applying the semigroup property of the transition amplitude,
\begin{equation}\label{semigroup}
K((n+1)\e,0) = \int\rmd\mu(\n_n)\int\rmd\mu(\{\z_\bk\})
K((n+1)\e,n\e) K(n\e,0)
\end{equation}
finite difference relations can be found for the coefficients in
the action. Setting $h=1$, and absorbing factors of $2\pi$, they are,
\begin{equation} \label{inductive1}
\begin{array}{lll}
\psi_{n} = (1-i \w_o \e) \psi_{n-1} +\sum_{\bk} (i \la_{n,\bk} \e) \phi_{n-1,\bk} && \psi_0 =\n_i \\
\phi_{n,\bk} =  (i \lab_{n,k} \e) \psi_{n-1} + (1-i \w_{\bk} \e)
\phi_{n-1,\bk} && \phi_{0,\bk} =0
\end{array}
\end{equation}
\begin{equation} \label{inductive2}
\begin{array}{lll}
g_{n,\bk} = (1-i \w_o \e) g_{n-1,\bk} + (i \la_{n,\bk} \e) f_{n-1,\bk} && g_{0,\bk} = 0 \\
f_{n,\bk} =  (i \lab_{n,\bk} \e) \sum_{{\bf l}} g_{n-1,{\bf l}}
+(1-i \w_{\bk} \e) f_{n-1,\bk} && f_{0,\bk} = z_{i, \bk}.
\end{array}
\end{equation}
The coupling constants in Eqs.~(\ref{inductive1}-\ref{inductive2})
have time indices because they are separated by complete sets of
states at different time steps when the Hamiltonian is partitioned,
thus they are separate sets of
Grassmann pairs. The transition amplitude at time $t$($=N\epsilon$) can be
written:
\begin{equation}\label{general amplitude}
K(t,0) = \exp \{ \nb_f \psi(t) +\sum_k \zb_{fk} f_{k}(t) +\sum_k
\nb_f g_{k}(t) +\sum_k \zb_{fk} \phi_{k}(t) \}.
\end{equation}
Since this equation is a function of Grassmann variables it is to
be treated as a formal expression that has meaning only in its
polynomial expansion. In that polynomial expansion many terms will
be truncated due to the nilpotency of the Grassmann variables.
Expanding out Eq.~(\ref{general amplitude}) and defining the
functionals
\begin{eqnarray}
\label{functional def1} F[\{m_\xi\}](t) &=& \prod_k (f_{k}(t))^{m_k} \\
\label{functional def2} G_{l}[\{m_\xi\}](t) &=& g_{l}(t) \prod_k (f_{k}(t))^{m_k} \\
\label{functional def3} \Psif[\{m_\xi\}](t) &=& \psi(t) \prod_k (f_{k}(t))^{m_k} \\
\label{functional def4} \Phif_p[\{m_\xi\}](t) &=& \phi_{p}(t) \prod_k (f_{k}(t))^{m_k} \\
\label{functional def5} \Phig_{lp}[\{m_\xi\}](t) &=& g_{p}(t) \phi_{l}(t)
\prod_k (f_{k}(t))^{m_k} \\
\label{functional def6} \Psig_{p}[\{m_\xi\}](t) &=& g_{p}(t) \psi(t) \prod_k (f_{k}(t))^{m_k}
\end{eqnarray}
gives the following expanded expression for the transition
amplitude (with time dependence left implied for notational clarity)
\begin{equation}\label{expanded amplitude}
\begin{split}
K(t,0) =\sum_{\{m_\xi\}} \left[\prod_\bk \frac{(\zb_{f\bk})^{m_\bk}}{m_k !}\right] \bigg( F[\{m_\xi\}] &+\nb_f \Psif [\{m_\xi\}] +\sum_l \nb_f G_{l}[\{m_\xi\}] \\
&+\sum_p \zb_{fp} \Phif_p[\{m_\xi\}] +\sum_{pl} \nb_f \zb_{fp}
\Phig_{lp}[\{m_\xi\}] \bigg)
\end{split}
\end{equation}
The variable $m_\xi$ is the number of photons in the $\xi^{th}$
mode of the final EMF state. The transition amplitude as written above
is a functional
sum over all distributions $\{m_\xi\}$. Differential equations for
the functionals that appear in the transition amplitude can be
found from the finite difference equations of
Eqs.~(\ref{inductive1}-\ref{inductive2}).
\begin{eqnarray}
\label{functional1} \dF[\{m_\xi\}] &=& -i \sum_q m_q \w_q F[\{m_\xi\}] + i \sum_{lp} G_p[\{m_\xi-\delta_{\xi l}\}] \\
\label{functional2} \dG_p[\{m_\xi\}] &=& -i(\w_o +\sum_q m_q \w_q) G_p[\{m_\xi\}] + i \la_p F[\{m_\xi+ \delta_{\xi p}\}] \nonumber\\ \\
\label{functional3} \dPsif[\{m_\xi\}] &=& -i(\w_o +\sum_q m_q \w_q) \Psif[\{m_\xi\}] + i \sum_p \la_p \Phif_p[\{m_\xi\}] \nonumber\\
&\mbox{ }&\mbox{ }\mbox{ }\mbox{ }\mbox{ }\mbox{ }+i \sum_{lp} m_l \la_l \Psig_p[\{m_\xi-\delta_{\xi l}\}] \\
\label{functional4} \dPsig_p[\{m_\xi\}] &=& -i(2\w_o + \sum_q m_q \w_q) \Psig_p[\{m_\xi\}] -i \sum_l \la_l \Phig_{lp}[\{m_\xi\}]  \nonumber\\
&\mbox{ }&\mbox{ }\mbox{ }\mbox{ }\mbox{ }\mbox{ }+i \la_p \Psif[\{m_\xi+\delta_{\xi p}\}] \\
\label{functional5} \dPhif_p[\{m_\xi\}] &=& -i(\w_p +\sum_q m_q \w_q) \Phif_p[\{m_\xi\}] +i \la_p \Psif[\{m_\xi\}]  \nonumber\\
&\mbox{ }&\mbox{ }\mbox{ }\mbox{ }\mbox{ }\mbox{ }+i\sum_{ql} m_q \la_q \Phig_{pl}[\{m_\xi-\delta_{\xi q}\}] \\
\label{functional6} \dPhig_{lp}[\{m_\xi\}] &=& -i(\w_o +\w_l +\sum_q m_q \w_q) \Phig_{lp}[\{m_\xi\}] -i\la_l \Psig_p[\{m_\xi\}]  \nonumber\\
&\mbox{ }&\mbox{ }\mbox{ }\mbox{ }\mbox{ }\mbox{ }+i\la_p
\Phif_l[\{m_\xi+\delta_{\xi p}\}]
\end{eqnarray}
Although the transition amplitude of Eq.~(\ref{expanded
amplitude}) and the differential equations with all Grassmann
variables removed of Eqs.~(\ref{functional1}-\ref{functional6})
can be used from this point onward, it is simpler instead to work
with Eq.~(\ref{general amplitude}) during the trace over final EMF
states. In the next section we shall combine forward and backward
versions of the transition amplitude to construct the reduced
propagator.

\subsection{Reduced Propagator}
The evolution of the reduced system with an initial atomic state
is given by,
\begin{equation} \label{evolved1}
\begin{split}
\rho(t) =
\int d\mu(\n_i)d\mu(\np_i) \prod_k \left[
d\mu(z_{ik})d\mu(\zp_{ik}) \right] J_R(t,0) R(0),
\end{split}
\end{equation}
in which $R(0)$ is the combined initial system-bath density
operator and $J_R(t,0)$ is the propagator for the reduced system,
\begin{equation}
J_R(t,0) = \int d\mu(\{\z_{fk}\}) K(t,0) \Kbp(t,0).
\end{equation}
Carrying out the integration with Eq.~(\ref{general amplitude})
and its barred conjugate one finds,
\begin{equation}\label{reduced propagator}
\begin{split}
J_R(t,0) = \exp \{ \nb_f \psi(t) +\psibp(t) \np_f &+\sum_\bk \nb_f \g_{\bk}(t) +\sum_\bk \gbp_{\bk}(t) \np_f \\
&+\sum_k \left(\fbp_{k}(t) +\phibp_{k}(t) \right) \left(f_{k}(t)
+\phi_{k}(t) \right) \}
\end{split}
\end{equation}

\subsection{Initial Thermal State}
For thermal vacuum the initial state in the coherent state representation and in units such that Boltzmann's constant is unity ($\mbox{k}_b=1$) is,
\begin{equation}\label{initial thermal state}
\begin{split}
R(0) = \left[ \prod_k \exp\{ e^{-\beta \w_k} \zb_{ik} \zp_{ik} \} \right] \times [\rho_{00}
+ \nb_i \rho_{10} + \np_i \rho_{01} + \nb_i \np_i \rho_{11}]
\end{split}
\end{equation}
Evaluating Eq.~(\ref{evolved1}) with substitutions from
Eq.~(\ref{reduced propagator}) and Eq.~(\ref{initial thermal
state}) one may obtain the evolved reduced density operator. After
expanding completely, the reduced density matrix elements become, for the upper state population,
\begin{equation}\label{rdm element 11}
\begin{split}
\rho_{11}(t) =
\rho_{00} \sum_{\{m_\xi\}} &\sum_l m_l \G_l[\{m_\xi-\delta_{\xi l}\}] \Gbp_l[\{m_\xi-\delta_{\xi l}\}] \mbox{ }e^{-\beta\sum_q m_q \omega_q} \\
+\rho_{11} \sum_{\{m_\xi\}} &\left( \Psif[\{m_\xi\}] +\sum_l m_l \Phig_{ll}[\{m_\xi-\delta_{\xi l}\}] \right) \\
&\times\left( \Psifbp[\{m_\xi\}] +\sum_l m_l
\Phigbp_{ll}[\{m_\xi-\delta_{\xi l}\}] \right) e^{-\beta\sum_q m_q \omega_q},
\end{split}
\end{equation}
for the lower state population,
\begin{equation}\label{rdm element 00}
\begin{split}
\rho_{00}(t) = \rho_{11} \sum_{\{m_\xi\}} &\sum_l (m_l +1) \Phif_{l}[\{m_\xi\}] \Phifbp_{l}[\{m_\xi\}] \mbox{ }e^{-\beta\sum_q m_q \omega_q} \\
&+\rho_{00} \sum_{\{m_\xi\}} \F[\{m_\xi\}] \Fbp[\{m_\xi\}] \mbox{ }e^{-\beta\sum_q m_q \omega_q},
\end{split}
\end{equation}
and for the off-diagonal,
\begin{equation}\label{rdm element 10}
\begin{split}
\rho_{10}(t) =
\rho_{10} \sum_{\{m_k\}} &\left( \Psif[\{m_\xi\}]+\sum_l m_l
\Phig_{ll}[\{m_\xi-\delta_{\xi l}\}] \right) \Fbp[\{m_\xi\}] \mbox{
}e^{-\beta\sum_q m_q \omega_q}
\end{split}
\end{equation}
in terms of the definitions of Eqs.~(\ref{functional
def1}-\ref{functional def6}), with $\{\rho_{11},\rho_{10},\rho_{01},\rho_{00}\}$ being the initial qubit density matrix elements.

\subsubsection{Low temperature}

The computation of the reduced density matrix elements involves
the calculation of the  functionals of Eqs.~(\ref{functional
def1}-\ref{functional def6}) and the evaluation of the functional
summations in Eqs.~(\ref{rdm element 11}-\ref{rdm element 10}). In
order to calculate the functionals, a low temperature and a weak
coupling approximation are applied to
Eqs.~(\ref{functional1}-\ref{functional6}). Details of the calculation
are shown in Appendix A. The resulting
expressions for the reduced density matrix elements, valid at low
temperature ($\rme^{-\beta\omega_o} <<1$) and weak coupling ($\lambda^2 <<1$), are
\begin{eqnarray}
\label{low temperature rdm 11} \rho_{11}(t) &=& \bigg[1- \Upsilon(t)\bigg]\rho_{00} +\bigg[1-\bigg(\frac{1-e^{-\Gamma_o t}}{1-e^{-\beta\omega_o -\Gamma_o t}}\bigg) \Upsilon(t) \bigg]\rho_{11} \\
\label{low temperature rdm 00} \rho_{00}(t) &=& \Upsilon(t) \rho_{00} +\bigg(\frac{1-e^{-\Gamma_o t}}{1-e^{-\beta\omega_o -\Gamma_o t}}\bigg) \Upsilon(t) \rho_{11} \\
\label{low temperature rdm 10} \rho_{10}(t) &=& {\rm e}^{-\Gamma_o
t/2 -{\rm i}\w_o t} \Upsilon(t) \rho_{10}
\end{eqnarray}
with the definition
\begin{equation} \label{upsilon}
\Upsilon(t) = \frac{1-e^{-\beta\omega_o}}{1-e^{-\beta\omega_o
-\Gamma_o t}}
\end{equation}
and $\Gamma_o = \frac{2\la^2 \w_o}{\pi}$ being the zero
temperature spontaneous emission rate. These reduced density matrix elements are illustrated in Fig.~(1).

\begin{figure}[t]
\includegraphics[width=8cm]{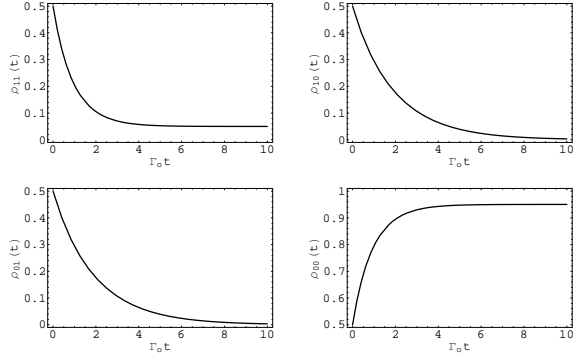}
\caption[Matrix Elements]{These plots illustrate the
non-Markovian reduced qubit matrix elements from Eqs.~(\ref{low
temperature rdm 11}-\ref{upsilon}) for the case of an initial
$\sigma_x$ eigenstate, $(|0\rangle+ |1\rangle)/\sqrt{2}$ at low
temperature ($\rme^{-\beta\omega_o}=0.05$), versus a
dimensionless time in units of $\Gamma_o^{-1}$ where $\Gamma_o$
is the zero temperature spontaneous emission rate. The diagonal
matrix elements thermalize to the low temperature values of
Eqs.~(\ref{rho11asymptote}-\ref{rho00asymptote}), and the
off-diagonal matrix elements decohere non-exponentially.}
\end{figure}

In the long time limit the populations tend to the following
thermal values valid at low temperature,
\begin{eqnarray}
\label{rho11asymptote}\rho_{11}(t\rightarrow\infty) &=& {\rm e}^{-\beta\w_o} \\
\label{rho00asymptote}\rho_{00}(t\rightarrow\infty) &=& 1 -{\rm e}^{-\beta\w_o}
\end{eqnarray}
and the off-diagonal coherence decays completely
\begin{equation}
\label{rho10asymptote}\rho_{10}(t\rightarrow\infty) = 0.
\end{equation}

\subsubsection{Zero temperature limit}
At zero temperature $\beta=\infty$ and Eqs.~(\ref{low temperature
rdm 11}-\ref{low temperature rdm 10}) become,
\begin{eqnarray}
\label{zero temperature rdm 11} \rho_{11}(t) &=& \rho_{11} e^{-\Gamma_o t} \\
\label{zero temperature rdm 00} \rho_{00}(t) &=& \rho_{00} +\rho_{11} \bigg(1-e^{-\Gamma_o t}\bigg) \\
\label{zero temperature rdm 10} \rho_{10}(t) &=& \rho_{10} {\rm
e}^{-\Gamma_o t/2 -{\rm i}\w_o t}
\end{eqnarray}
which is the expected result from Ref.~\cite{ABH}.

\subsubsection{The Markov approximation limit}
For reference purposes we also include the results in the Markov
approximation, which are  valid in the regime of high temperature which are,
\begin{eqnarray}
\rho_{11}(t) &=& \rho_{11}(0) \rme^{-\Gamma_o \coth(\beta\omega_o
/2) t} +
\frac{\rme^{-\beta\omega_0}}{1+\rme^{-\beta\omega_0}} (1-\rme^{-\Gamma_o
\coth(\beta\omega_o /2) t})\\
\rho_{00}(t) &=& 1- \rho_{11}(t)\\
\rho_{10}(t) &=& \rho_{10}(0) \rme^{-i\omega_o t-\frac{\Gamma_o
t}{2} \coth(\beta\omega_o /2)}.
\end{eqnarray}
Their asymptotic values are,
\begin{eqnarray}
\label{rhom11asymptote}\rho_{11}(t\rightarrow\infty) &=& \frac{\rme^{-\beta\w_o}}{1+\rme^{-\beta\w_o}} \approx \rme^{-\beta\w_o} +O((\rme^{-\beta\w_o})^2) \\
\label{rhom00asymptote}\rho_{00}(t\rightarrow\infty) &=& \frac{1}{1+\rme^{-\beta\w_o}} \approx 1 -\rme^{-\beta\w_o} +O((\rme^{-\beta\w_o})^2)
\end{eqnarray}
and the off-diagonal coherence decays completely
\begin{equation}
\label{rhom10asymptote}\rho_{10}(t\rightarrow\infty) = 0.
\end{equation}
The thermal populations in the non-Markovian low temperature approximation match the Markovian thermal populations up to $O((\rme^{-\beta\w_o})^2)$.

\section{Discussion}
\subsection{Decoherence}

\begin{figure}[t]
\includegraphics[width=8cm]{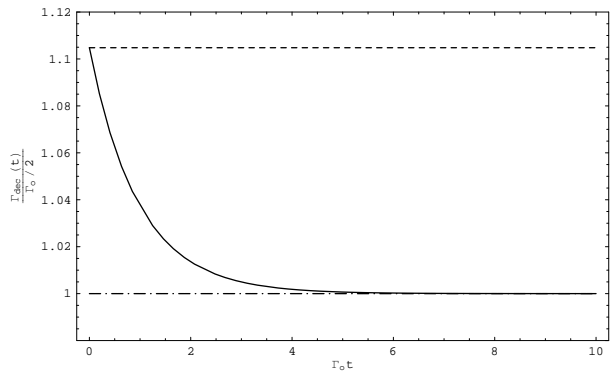}
\caption[Decoherence rate of a qubit in a low temperature
bath.]{This plot shows the quantity
$\frac{\Gamma_{dec}(t)}{\Gamma_o/2}$, (the ratio of the
non-Markovian decoherence rate defined by $\Gamma_{dec}(t) :=
-\frac{\dot{\rho}_{10}(t)}{\rho_{10}(t)}$, over the zero
temperature decoherence rate, $\Gamma_o/2$) versus the
dimensionless time $\Gamma_o t$ in the low temperature regime,
$e^{-\beta\w_o} =0.05$. The dotted line is the value of the
Markovian prediction at finite temperature, $\coth(\beta\w_o/2)$.
Initially the non-Markovian prediction matches the Markovian
result (at the dotted line). As the qubit and EMF become
correlated the reduced dynamics deviates from the Markovian
prediction and the decoherence rate asymptotes to the zero
temperature value (along the dashed-dotted line).}
\end{figure}

The decoherence rate is found by computing the off-diagonal
elements of the reduced density matrix (e.g.~$\rho_{10}(t)$). The
inclusion of bath as well as system dynamics causes the fall off
of the off-diagonal matrix elements to become slightly
sub-exponential. From previous work~\cite{ABH} we know that at
zero temperature the decoherence rate is $\Gamma_0/2 = \lambda^2
\omega_o / \pi$. Markovian approaches (e.g.~\cite{carmichael})
predict a decoherence rate of $\Gamma_0 \coth(\beta\omega_o/2)/2
$, valid at high temperatures. The present calculation shows that
the decoherence rate, $\Gamma_{dec}(t) :=
-\frac{\dot{\rho}_{10}(t)}{\rho_{10}(t)}$, actually changes as
the total system evolves. As shown in Fig.~(2) the decoherence
rate at $t=0$, when the bath is by assumption in a thermal state
uncorrelated with the qubit, agrees with the prediction of
Markovian approaches. As the system and bath evolve together the
decoherence rate falls back down to the zero temperature value.
The interpretation of this is: initially the two cases have the
same decoherence rate because by arrangement the combined system
is a product state of qubit and thermal bath, which is the state
assumed in Markovian  approaches (there is no prior correlation).
As the system and bath interact, the correlations that arise
alter the reduced system dynamics and the combined state evolves
away from that initial factorizable state. The overall effect is
that the the qubit decoheres more slowly in non-Markovian dynamics
than in Markovian dynamics.

\subsection{Relaxation}

\begin{figure}[t]
\includegraphics[width=8cm]{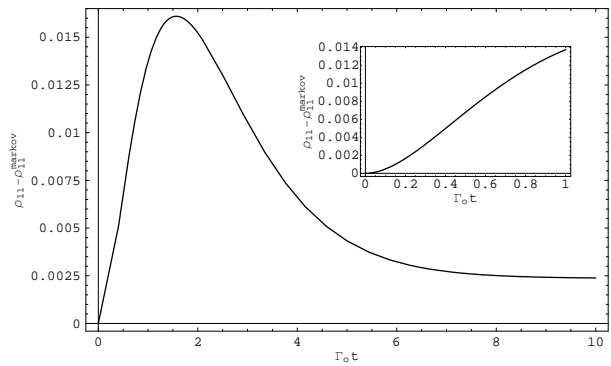}
\caption[Relaxation of a qubit in a low temperature bath]{This
plot shows the difference in the upper state populations,
$\rho_{11}(t) -\rho^{\mbox{markov}}_{11}(t)$, between the
non-Markovian prediction, $\rho_{11}(t)$, and the Markovian
prediction, $\rho^{\mbox{markov}}_{11}(t)$, given that the qubit
is initially in the upper state (i.e.~$\rho_{11}(0)=1$). The
difference is plotted versus dimensionless time $\Gamma_o t$ and
in the low temperature regime, $e^{-\beta\w_o}=0.05$. Inspection
of the plot shows that in non-Markovian dynamics the upper state
decays more slowly than in Markovian dynamics. At long times the
difference in the populations is zero up to
$O((\rme^{-\beta\w_o})^2)$ (see Eq.~(\ref{rhom11asymptote})). The
inset shows that the non-Markovian and Markovian predictions
agree initially.}
\end{figure}

The relaxation time scale is measured by the value of
$\rho_{11}(t)$, assuming that $\rho_{11}(0) = 1$. Similar to the
case of decoherence, because the initial state of the combined
system-bath is taken to be a product state of qubit and thermal
bath, as it is in Markovian approaches, the dynamics for the populations initially agree in non-Markovian and Markovian dynamics (see the inset of Fig.~(3)). Then as the system and bath interact, the non-Markovian result for
the dynamics of the reduced system, which takes into consideration
the dynamics of both the bath and the qubit, deviates from the
Markovian prediction, as shown in Fig.~(3). However, the long time
behavior of our prediction matches the thermalization prediction
of the Markovian prediction up to $O((\rme^{-\beta\w_o})^2)$. Most importantly, Fig.~(3) shows that the upper state population relaxes more slowly in non-Markovian dynamics than in Markovian dynamics.

\begin{figure}[t]
\includegraphics[width=8cm]{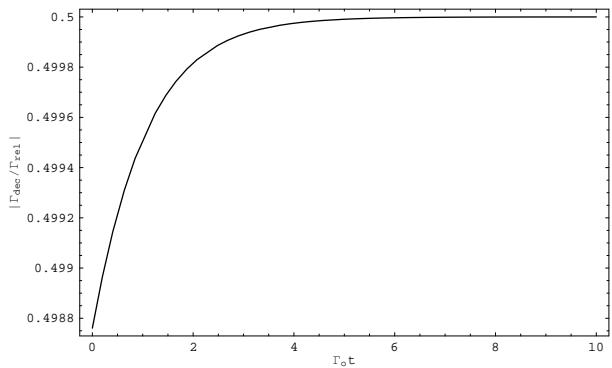}
\caption[Ratio of decoherence to relaxation rates]{This plot shows
the ratio of the non-Markovian decoherence rate to the
non-Markovian relaxation rate as a function of dimensionless
time, $\Gamma_o t$, in the low temperature regime, $e^{-\beta\w_o}
=0.05$. The value is approximately constant at~0.5, which is also
the ratio between the Markovian decoherence and relaxation rates.
The is because both cases share the same physical determining
factor, i.e.~that the resonant type of coupling is at work in
this model.}
\end{figure}

We may define the relaxation rate (for the case that $\rho_{11}(0)
= 1$) as $\Gamma_{rel}(t) := -
\frac{\dot{\rho}_{11}(t)}{\rho_{11}(t) - \rho_{11}(\infty)}$. The
plot of Fig.~(4) shows the dependence of the ratio
$\Gamma_{dec}(t)/\Gamma_{rel}(t)$ on time. It demonstrates that
the relaxation and decoherence rate are of the same order of
magnitude. In other words, the rate of quantum phase information
escaping from the system to the environment is the same as the
rate of energy flow. This property is characteristic of the {\em
resonant} coupling between the two-level atom and the EMF, which
leads to a different decoherence behavior  from quantum Brownian
motion~(QBM) models~\cite{ABH}. One way to visualise the
distinction is the realization that in QBM the couplings allow the
interaction of the system with the far-infrared modes of the
environment. The system then loses the phase information through
soft photons  which however carry very little energy. Hence in QBM
systems,  the relaxation time is much longer than the decoherence
time. However, in resonant systems,  such as being studied here
and in~\cite{ABH},  the system interacts primarily with the modes
of the environment near the resonance frequency. Consequently, the
phase information escapes through photons of energy equal to that
of the atom and the decoherence rate is essentially the same with
relaxation rate. We should remark that although the present
results only hold for the low-temperature limit the near equality
of decoherence and relaxation rate is valid even in the high
temperature limit as can be seen already from the Markov
approximation.

\subsection{Entanglement}

\begin{figure}[t]
\includegraphics[width=8cm]{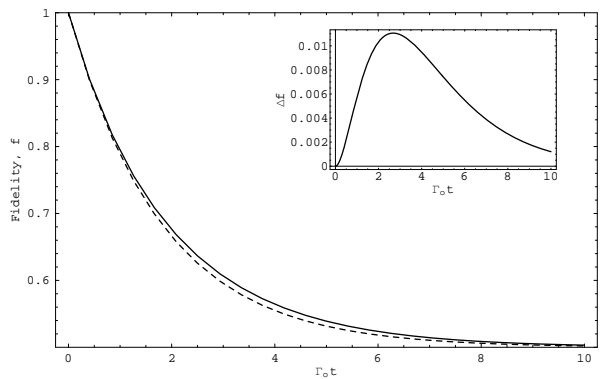}
\caption[Fidelity]{This plot shows the quantity
$\mbox{f}=\mbox{Tr}[\rho(t) U_o(t) \rho(0) U_o^\dagger(t)]$ as a
function of the dimensionless time, $\Gamma_o t$,  and at low
temperature ($e^{-\beta\w_o} =0.05$) for an initial $\sigma_x$
eigenstate, $(|0\rangle+|1\rangle)/\sqrt{2}$), with $U_o(t)$
being the free evolution operator. Being a measure of the
persistence of the initial qubit state after interaction with the
environment, it can be considered as the \textbf{fidelity} of the
qubit in its environment. The non-Markovian fidelity is plotted
as a solid line and the Markovian fidelity is plotted as a dashed
line. The inset is the non-Markovian fidelity minus the Markovian
fidelity. Inspection of it shows that in the non-Markovian
dynamics the EMF bath degrades the fidelity of the qubit more
slowly than in the Markovian dynamics.}
\end{figure}

\begin{figure}[t]
\includegraphics[width=8cm]{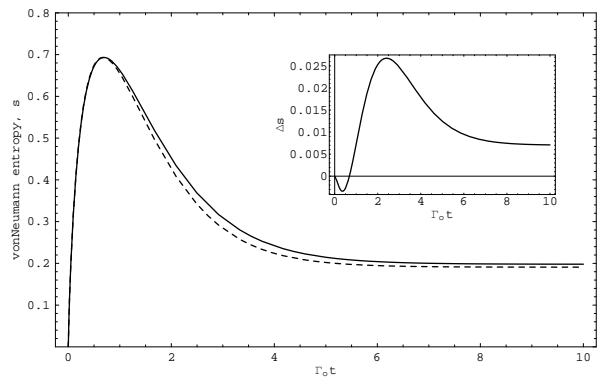}
\caption[vonNeumann entropy]{This plot shows the von~Neumann
entropy, $s(t)=-\mbox{Tr }\rho\log_e(\rho)$, of the reduced qubit
density matrix versus the dimensionless time, $\Gamma_o t$, for
low temperature ($e^{-\beta\w_o} =0.05$). The von~Neumann entropy
is a measure of the \textbf{purity} of a density operator. Both
the non-Markovian and Markovian von~Neumann entropies are
plotted, as solid and dashed lines, respectively. Inspection of
the plot shows that as the qubit interacts with the environment
it becomes more mixed, but as the qubit equilibrates with the
environment (after a time on the order of the relaxation time) it
becomes less mixed due to the low temperature. In the inset the
difference of the non-Markovian von~Neumann entropy minus the
Markovian von~Neumann entropy is plotted. It shows that during the
initial period of mixing non-Markovian dynamics predicts less
mixing than does Markovian dynamics. Then, during the later 
equilibration period Markovian dynamics predicts a less mixed
state. Finally, both dynamics reach a low temperature equilibrium 
state which is less mixed than at intermediate times. As in Fig.~(3), 
the non-Markovian dynamics asymptotes to thermal equilibrium more slowly 
than the Markovian dynamics.}
\end{figure}

There exists no computable measure of entanglement between a
qubit and an infinite continuous bath such as the electromagnetic
field. However, since Markovian predictions explicitly exclude
system-bath entanglement, comparison of those predictions with the
present results can reveal the effects of system-bath
entanglement. First is the \textbf{decoherence} rate discussed
previously (see Fig.~(2)). Its evolution from the thermal to the
zero temperature value shows that the combined system-bath
reaches and holds at some state in which the off-diagonal
elements of the qubit are no longer affected by the thermal
nature of the bath. No product state could give such behavior and
cause thermalization of the populations. Second is the
\textbf{fidelity}, $\mbox{f}=\mbox{Tr}[\rho(t) U_o(t) \rho(0)
U_o^\dagger(t)]$, of the $(|0\rangle + |1\rangle)/\sqrt{2}$ state
shown in Fig.~(5), with $U_o(t)$ being the free evolution
operator. The fidelity in this case is a measure of the
persistence of the initial qubit state after interaction with the
environment. Inspection of the difference between the
non-Markovian and Markovian predictions for fidelity (inset of
Fig.~(5)) shows that non-Markovian dynamics predicts a slower
loss of fidelity than Markovian dynamics, although with continued
interaction both dynamics predict complete loss of fidelity. Third
is the \textbf{von~Neumann entropy}, $s(t)=-\mbox{Tr
}\rho\log_e(\rho)$, for the initial qubit state $|1\rangle$,
shown in Fig.~(6). The von~Neumann entropy is a measure of the
\textbf{purity} of a density matrix. Inspection of the difference
in this case (inset of Fig.~(6)) shows that the Markovian result
initially predicts a greater loss of purity, but after a time on
the order of the decay timescale, it predicts less loss of purity
than the non-Markovian result. The reason for this seeming
contradiction is that at low temperature, the system in its
approach to thermal equilibrium, is driven  to a less mixed
state. Since correlations with the bath slow the drive to thermal
equilibrium in non-Markovian dynamics, as shown in Fig.~(3), this
process is slower in the non-Markovian regime. Comparison of
these three quantities (decoherence rate, fidelity and
von~Neumann entropy) shows a consistent picture in which
non-Markovian dynamics is characterized by the preservation of
coherence for longer time.

\subsection{Conclusion}
We have studied a two level atom coupled to an electromagnetic
field (EMF) at finite temperature in the multimode
Jaynes-Cummings model. We have computed the reduced evolution of
the two level system and addressed the issues of decoherence,
relaxation, and entanglement from its interaction with the EMF
bath. Our approach makes use of a modified influence functional
technique, which enables one to compute the reduced system
dynamics while including the entangled evolution of the bath and
qubit degrees of freedom. That is in contrast to Markovian
approaches, which assume a fixed bath and hence neglect any
dynamics in the bath. We adopt a Grassmann coherent state path
integral representation for the atom degrees of freedom and
bosonic coherent state path integral representation for the
electromagnetic field, and assume a weak coupling ($2^{nd}$~order
Born) approximation under low temperature conditions.

We find non-exponential decay for both the diagonal and
off-diagonal matrix elements of the qubit's reduced density
matrix. From the diagonal elements we see the qubit relax to
thermal equilibrium with the bath. From the off-diagonal elements,
we see the decoherence rate beginning at the rate usually
predicted for a thermal bath, but evolving to the zero
temperature decoherence rate as the qubit and bath become
entangled. Comparison of the relaxation and decoherence rates,
shown in Fig.~(4), reveals that as in the zero temperature case,
both rates are comparable. At short times the ratio of the
decoherence to the relaxation rate is initially smaller, but only
by a small amount. At higher temperatures the initial difference
between the two does increase, but that regime reaches the limits
of validity of the present results. We can see why at low
temperatures both rates are related to the atomic transition
rate, because it is the only relevant physical scale present
(unlike at finite temperature where the thermal scale is also at
work). This, in turn, is a consequence of the particular resonant
coupling between the two-level atom and the EMF, as explained
earlier in~\cite{ABH}.

On the entanglement of a qubit with the EMF (under resonant
coupling) we calculated the qubit's fidelity and the von~Neumann
entropy. The Markovian result predicts higher loss of fidelity
and purity as compared with the more accurate non-Markovian
calculations. Qualitatively, with the inclusion of quantum
correlations between the qubit and its environment, the
non-Markovian processes tend to slow down the drive of the system
to equilibrium, prolonging the decoherence and better preserving
the fidelity and purity of the system.\\

\noindent {\bf Acknowledgements} This work is supported in part by
ARDA contract MDA90401/C0903, a NSF and a NIST grant to the
University of Maryland.

\appendix
\section{Calculational details of qubit in a thermal bath}
\subsection{Approximated functional solutions}
Eqs.~(\ref{functional1}-\ref{functional6}) are two sets of coupled
differential equations. One set being the pair of equations
\begin{eqnarray}
\dot{F}[\{m_\xi\}] &=& -i \sum_q m_q \w_q F[\{m_\xi\}]
+ i \sum_{lp} m_l \lab_l G_p[\{m_\xi-\delta_{\xi l}\}] \\
\dot{G}_p[\{m_\xi\}] &=& -i(\w_o +\sum_q m_q \w_q) G_p[\{m_\xi\}]  + i
\la_p F[\{m_\xi+ \delta_{\xi p}\}]
\end{eqnarray}
and the remaining four equations comprising the other set. The
solution method for this pair in the low temperature and weak
coupling limits will be sketched out in this appendix. The
solutions for the other set in the same limits will follow a
similar sequence. First, given the initial conditions
\begin{eqnarray}
F [\{m_\xi\}] (t=0) &=& 1 \\
G _p[\{m_\xi\}] (t=0) &=& 0
\end{eqnarray}
the Laplace transforms of the above equations are
\begin{eqnarray}
\label{laplace functional1 1} z \tilde{F}[\{m_\xi\}](z) -1 &=& -i \sum_q
m_q \w_q \tilde{F}[\{m_\xi\}](z) +
i \sum_{lp} m_l \lab_l \tilde{G}_p[\{m_\xi-\delta_{\xi l}\}](z) \\
\label{laplace functional2 1} z \tilde{G}_p[\{m_\xi\}](z) &=& -i(\w_o
+\sum_q m_q \w_q) \tilde{G}_p[\{m_\xi\}](z) + i \la_p
\tilde{F}[\{m_\xi +\delta_{\xi p}\}](z).
\end{eqnarray}
The second equation can be rearranged into
\begin{equation}
\label{laplace functional2 2} \tilde{G}_p[\{m_\xi\}](z) = \frac{i \la_p
\tilde{F}[\{m_\xi+ \delta_{\xi p}\}](z)}{z +i(\w_o +\sum_q m_q \w_q)},
\end{equation}
which can be substituted back into Eq.~(\ref{laplace functional1
1}) to give
\begin{eqnarray}
\label{laplace functional1 2} \left(z +i \sum_q m_q \w_q
\right)\tilde{F}[\{m_\xi\}](z) = 1 + i \sum_{lp} \frac{i
m_l \lab_l \la_p \tilde{F}[\{m_\xi-\delta_{\xi l}+ \delta_{\xi p}\}](z)}{z
+i(\w_o-\w_l +\sum_q m_q \w_q)}.
\end{eqnarray}
In the expression above the  low temperature approximation is
applied by setting $p=l$ in the summation of the RHS. The
justification is that the summation on the RHS will be peaked
about $\w_l=\w_o$ such that the greatest contribution from
$\tilde{F}[\{m_\xi-\delta_{\xi l}+ \delta_{\xi p}\}](z)$ will be for
$\w_l=\w_o$. However, at low temperatures those frequencies will
not be populated. As a result the vacuum will be annihilated,
unless $\delta_{\xi p}=\delta_{\xi l}$, which will cause the major
contribution from the $p$ summation to be from $p=l$. The low temperature approximation is thus that the temperature is low enough that the modes with frequency $\omega_o$ are unoccupied, i.e. $\rme^{-\beta\omega_o} <<1$. Applying
this approximation, Eq.~(\ref{laplace functional1 2}) can be
rewritten as
\begin{eqnarray}
\label{laplace functional1 3} \tilde{F}[\{m_\xi\}](z) =  {}\left( z +i
\sum_q m_q \w_q   + \sum_{l} \frac{m_l \la_l^2}{z +i(\w_o-\w_l
+\sum_q m_q \w_q)} \right)^{-1}.
\end{eqnarray}
The zero$\mbox{}^{th}$  order pole of $\tilde{F}[\{m_\xi\}](z)$ is at
$z=-i \sum_q m_q \w_q$. The reaction term at this point is found equal
to $ \frac{\Gamma_o m_o}{2} +i\Delta$, with $\Gamma_o=\frac{\lambda^2 \omega_o}{\pi}$, which shows that the second
order shift in the pole includes both a real and an imaginary
part. After absorbing the imaginary part in a renormalization of
the frequency, the second order pole is $z=-i \sum_q m_q \w_q
-\frac{\Gamma_o m_o}{2}$ with the definitions
\begin{equation}
m_o = \sum_{\w_l=\w_o} m_l
\end{equation}
The desired functional can be calculated as in inverse Laplace
transform of Eq.~(\ref{laplace functional1 3}) at the second order
pole to give
\begin{equation}
\label{functional1 soln} F[\{m_\xi\}](t) = \exp \left\{-\frac{\Gamma_o
m_o}{2} t -i \sum_q m_q \w_q t  \right\}.
\end{equation}
The inverse Laplace transform contains a contribution of a branch cut as well as a pole~\cite{SekeHerfort88,DiVincenzoLoss04}. We ignore the contribution of the branch cut, which is negligible at all but very late times such that $\Gamma_o t >20$ and very early times such that $\Gamma_o t < 10^{-21}$ (see Eq.(3.20) of Ref.~\cite{SekeHerfort88}). In all cases, we assume that time is much later than the inverse cut-off
time. Further comparison between the branch cut and the non-Markovian correction over Markovian dynamics shows that the branch cut contribution is smaller by greater than three orders of magnitude for $\Gamma_o t > 0.1$. The other functional in  the pair can be calculated by integrating
Eq.~(\ref{functional2})
\begin{equation}
\label{functional2 soln} G_l[\{m_\xi -\delta_{\xi l}\}](t) =
i\frac{\la}{\sqrt{\w_l}} \frac{1- e^{-\frac{\Gamma_o m_o}{2} t -i
\sum_q m_q \w_q t}}{\frac{\Gamma_o m_o}{2} +i(\w_l -\w_o)} \mbox{ }
e^{i(\w_l -\w_o -\sum_q m_q \w_q)t}.
\end{equation}
Following similar calculations the rest of the functionals are
found to be
\begin{align}
\label{functional3 soln} &\Psif[\{m_\xi\}] (t) = e^{ -\frac{\Gamma_o (m_o +1)}{2}t -i (\w_o +\sum_q m_q \w_q)t } \\
\label{functional4 soln} &\Psig_p[\{m_\xi -\delta_{\xi p}\}] (t) =
\frac{\la e^{-\frac{\Gamma_o}{2}t -i(\w_o +
\sum_q m_q \w_q)t }}{\sqrt{\w_p}(\w_p -\w_o -i\frac{\Gamma_o m_o}{2})} \left[ e^{i(\w_p -\w_o)t} -e^{-\frac{\Gamma_o m_o}{2}t} \right] \\
\label{functional5 soln} &\Phif_p[\{m_\xi\}] (t) = \frac{\la
e^{-\frac{\Gamma_o m_o}{2}t -i(\w_o +\sum_q m_q \w_q)t}}
{\sqrt{\w_p}(\w_p -\w_o -i\frac{\Gamma_o}{2})}  \left[e^{-\frac{\Gamma_o}{2}t} -e^{i(\w_p -\w_o)t} \right] \\
\label{functional6 soln} &\Phig_{lp}[\{m_\xi -\delta_{\xi p}\}] (t) =
\frac{\la^2 e^{-i(\w_o +\w_l -\w_p +
\sum_q m_q \w_q)t}}{\sqrt{\w_l}\sqrt{\w_p}}
\nonumber \\ &{}\hspace{4cm} \times
\frac{\left(e^{ -\frac{\Gamma_o}{2}t + i(\w_l -\w_o)t }
-1 \right) \left(1 - e^{ -\frac{\Gamma_o m_o}{2}t -i(\w_p
-\w_o)t}\right) }{\left[ (\w_l -\w_o) +i\frac{\Gamma_o}{2} \right]
\left[  (\w_p -\w_o) -i\frac{\Gamma_o m_o}{2}\right]}
\end{align}

\subsection{Computation of density matrix elements}
The solutions of Eqs.~(\ref{functional1 soln}-\ref{functional6
soln}) can be substituted into Eqs.~(\ref{rdm element 11}-\ref{rdm
element 10}) to evaluate the reduced density matrix elements in
the limits of low temperature and weak coupling. The reduced
density matrix elements in that form are summations over all
distributions $\{m_\xi\}$. The $\rho_{10}(t)$ matrix element will be
demonstrated below as a representative calculation. The evaluation
of the other summations follow along similar lines. From
Eq.~(\ref{rdm element 10}), the off-diagonal density matrix
element is
\begin{equation}
\begin{split}
\rho_{10}(t) = \rho_{10} \sum_{\{m_\xi\}} \left( \Psif[\{m_\xi\}]+\sum_l
m_l \Phig_{ll}[\{m_\xi-\delta_{\xi l}\}] \right)
\Fbp[\{m_\xi\}] \mbox{ }e^{-\beta\sum_q m_q \omega_q}.
\end{split}
\end{equation}
First, from Eqs.~(\ref{functional1 soln}-\ref{functional6 soln})
the functional in parentheses can be determined to be
\begin{equation}
\Psif[\{m_\xi\}]+\sum_l m_l \Phig_{ll}[\{m_\xi-\delta_{\xi l}\}] = e^{
-\frac{\Gamma_o (m_o +1)}{2}t -i(\w_o +\sum_q m_q \w_q)t},
\end{equation}
so that the off-diagonal matrix element becomes
\begin{equation}
\rho_{10}(t) = \rho_{10} \sum_{\{m_\xi\}} \exp\left\{ -\frac{\Gamma_o
(2 m_o +1)}{2}t -i \w_o t \right\} \mbox{ }e^{-\beta\sum_q m_q \omega_q}.
\end{equation}
Denoting by primes those terms for which $\w_\xi=\w_o$ and double
primes those for which $\w_\xi\neq\w_o$, the summand can be
rewritten with the substitution $m_o =\sum_\xi^\prime m_\xi$,
\begin{align}
\label{rdm element 10 2} \rho_{10}(t) =  \rho_{10}\mbox{
}e^{-\frac{\Gamma_o}{2}t -i \w_o t} \sum_{\{m_\xi\}} \prod_\xi^\prime e^{
-(\Gamma_o t +\beta\w_o) m_\xi} \prod_\xi^{\prime\prime} e^{ -\beta\w_\xi
m_\xi}.
\end{align}
The summation over distributions can be more clearly written as
\begin{equation}
\sum_{\{m_\xi\}} = \left[\prod_\xi \sum_{m_\xi=0}^\infty \right] =
\left[\prod_\xi^\prime \sum_{m_\xi=0}^\infty \right]
\left[\prod_\xi^{\prime\prime} \sum_{m_\xi=0}^\infty \right],
\end{equation}
which allows us to bring Eq.~(\ref{rdm element 10 2}) into the
form
\begin{eqnarray}
\rho_{10}(t) = \rho_{10} \mbox{ }e^{-\frac{\Gamma_o}{2}t -i \w_o t} \left(
\frac{1- e^{-\beta\w_o}}{1- e^{-(\Gamma_o t +\beta\w_o)}} \right)
  e^{ -\sum_q \ln\left[1- e^{-\beta\w_q}\right]}
\end{eqnarray}
The factor at the end is removed by normalization of the reduced
matrix element by its value if $\Gamma_o=0$. The final result for the
off-diagonal matrix element is
\begin{eqnarray}
\rho_{10}(t) = \rho_{10}\mbox{ }e^{-\frac{\Gamma_o}{2}t -i \w_o t}
\left( \frac{1- e^{-\beta\w_o}}{1- e^{-(\Gamma_o t +\beta\w_o)}}
\right)
\end{eqnarray}
with $\Gamma_o$ being the zero temperature spontaneous emission rate. The other
reduced density matrix elements are given in the text.

\end{document}